\documentclass[a4paper,11pt]{article}
\usepackage{aas_macros}
\usepackage{pos}

\title{SNAD: enabling discovery in the era of big data}

\author*[a,b]{Maria Pruzhinskaya}
\author[b]{Emille E. O. Ishida}
\author[c]{Konstantin Malanchev}
\author[a]{Anastasia Lavrukhina}
\author[b,d]{Etienne Russeil}
\author[a,e]{Timofey Semenikhin}
\author[f]{Sreevarsha Sreejith}
\author[b]{Emmanuel Gangler}
\author[a,g]{Matwey Kornilov}
\author[]{Vladimir Korolev}
\author[a,h]{Alina Volnova}

\affiliation[a]{Lomonosov Moscow State University, Sternberg Astronomical Institute,\\
  Universitetsky pr.~13, Moscow, 119234, Russia}

\affiliation[b]{Universit\'e Clermont Auvergne, CNRS, LPCA, \\ 63000 Clermont-Ferrand, France}

\affiliation[c]{McWilliams Center for Cosmology \& Astrophysics, Department of Physics, Carnegie Mellon University, \\ Pittsburgh, PA 15213, USA}

\affiliation[d]{The Oskar Klein Centre, Department of Astronomy, Stockholm University, \\AlbaNova, SE-10691 Stockholm, Sweden}

\affiliation[e]{Lomonosov Moscow State University, Faculty of Physics,\\ Leninsky Gori 1, Moscow 119234, Russia}

\affiliation[f]{Physics department, University of Surrey, \\ Stag Hill, Guildford GU2 7XH}

\affiliation[g]{National Research University Higher School of Economics, \\ 21/4 Staraya Basmannaya Ulitsa, Moscow, 105066, Russia}

\affiliation[h]{Space Research Institute of the Russian Academy of Sciences (IKI),\\ 84/32 Profsoyuznaya Street, Moscow, 117997, Russia}

\emailAdd{pruzhinskaya@gmail.com}

\abstract{In the era of wide-field surveys and big data in astronomy, the SNAD team (\url{https://snad.space}) is exploiting the potential of modern datasets for discovering new, unforeseen, or rare astrophysical objects and phenomena with machine learning (ML). The SNAD pipeline was built under the hypothesis that, although automatic ML algorithms have a crucial role to play in this task, the scientific discovery is only completely realized when such systems are designed to boost the impact of domain knowledge experts. Our key contributions include the development of the \texttt{Coniferest} Python library, which offers implementations of two active learning algorithms with an “expert in loop”, and the creation of the SNAD Transient Miner, facilitating the search for specific types of transients. We have also developed the SNAD Viewer, a web portal that provides a centralized view of individual objects from the Zwicky Transient Facility’s (ZTF) data releases, making the analysis of potential anomalies more efficient. Finally, when applied to ZTF data, our approach has resulted in more than a hundred new supernova (SN) candidates, along with a few other non-catalogued objects, such as red dwarf flares, superluminous SNe, RS CVn type variables, and young stellar objects.}

\FullConference{Frontier Research in Astrophysics – IV (FRAPWS2024)\\
 9-14 September 2024\\
Mondello, Palermo, Italy\\}

\begin{document}
\maketitle

\section{Introduction}

Unexpected discoveries have played a central role in the history of astronomy. 
Many important phenomena were not found through targeted searches, but were discovered unexpectedly during observations originally designed for other purposes. 
Classical examples include the discovery of the cosmic microwave background radiation by Arno Penzias and Robert Wilson \citep{1965ApJ...142..419P}, the discovery of gamma-ray bursts by the Vela satellites \citep{1973ApJ...182L..85K}, and the discovery of pulsars in 1967 by Jocelyn Bell and Antony Hewish \citep{1968Natur.217..709H}. Even the discovery of the accelerating expansion of the Universe at the end of the 1990s may be regarded, in some sense, as an unexpected result: distant SNe~Ia were initially observed in order to measure the deceleration of the cosmic expansion \citep{1997ApJ...483..565P,1998ApJ...493L..53G}, but instead led to the opposite conclusion \citep{1998AJ....116.1009R,1999ApJ...517..565P}.

This historical picture has changed profoundly with the arrival of large-scale sky surveys. Beginning with projects such as the SDSS, whose imaging catalogue contains photometry for almost half a billion unique objects \citep{2017AJ....154...28B,2022ApJS..259...35A}, and more recently DESI, whose first public data release includes nearly 20 million spectra \citep{2019BAAS...51g..57L,2026AJ....171..285D}, and continuing with time-domain facilities such as the Zwicky Transient Facility (ZTF, \cite{Bellm_2019}) and Rubin Legacy Survey of Space and Time (LSST, \cite{LSST2009}), astronomy has entered an era of massive and complex data sets. 
In such conditions, the serendipitous discovery of rare or previously unknown astrophysical phenomena becomes increasingly unlikely. At the same time, it is remarkable that manual screening continues to play a role in some discoveries, illustrating the scientific value of human inspection, e.g. the discovery of the first interstellar comet, 2I/Borisov \citep{2019CBET.4666....1B} or the identification of compact, extremely star-forming galaxies, known as ``Green Peas'', first noted by volunteers in the Galaxy Zoo project \citep{2009MNRAS.399.1191C}.

To make such discoveries feasible in the modern era of big astronomical data, the use of machine-learning methods has become unavoidable. Over the last decade, the astronomical community has devoted considerable effort to this problem, applying a broad variety of anomaly-detection (AD) strategies to different kinds of data, including images, spectra, and light curves. Some representative examples are listed below.

For changing-state active galactic nuclei,  Sánchez-Sáez et al. \cite{2021AJ....162..206S} identified 75 promising candidates by applying deep-learning-based dimensionality reduction combined with Isolation Forest to ZTF~DR5 light curves. In imaging data, Storey-Fisher et al. \cite{fisher2021} applied a Wasserstein generative adversarial network to nearly one million optical galaxy images from the Hyper Suprime-Cam survey, identifying anomalies of interest such as galaxy mergers, tidal features, and extreme star-forming galaxies. More recently, O’Ryan \& Gómez \cite{2025A&A...704A.227O} used the \texttt{AnomalyMatch} method, which combines semi-supervised and active learning techniques, on about 100 million image cutouts from the Hubble Legacy Archive and identified 86 new candidate gravitational lenses, 18 jellyfish galaxies, and 417 mergers or interacting galaxies.

Spectroscopic data have also been explored in this context. Using an unsupervised Random Forest on more than two million galaxy spectra from the SDSS, Baron \& Poznanski \cite{2017MNRAS.465.4530B} examined the 400 objects with the highest outlier scores and identified galaxies with extreme emission-line ratios, abnormally strong absorption features, unusual continua, close galaxy pairs, and galaxies that host supernovae.

Variable stars and transients provide another application for anomaly detection. Martínez-Galarza et al. \cite{martinez2021} combined tree-based anomaly-detection methods with manifold learning to identify unusual light curves in Kepler data. Giles \& Walkowicz \cite{2019MNRAS.484..834G} applied a clustering method
to a set of features derived for Kepler light curves and
demonstrated that their method can recover anomalies such as Boyajian’s star, as well as cataclysmic variables. Sarkar et al. \cite{sarkar2022} used the Earth as an anomaly example in order to estimate the habitability of exoplanets using a multi-stage memetic algorithm. 

Rebbapragada et al. \cite{2009arXiv0905.3428R} introduced PCAD, an unsupervised anomaly-detection method designed specifically for large sets of unsynchronized periodic time-series data, and found among its top-ranked anomalies a mixture of noisy light curves, misclassified variables, and several atypical Cepheids and eclipsing binaries in the Optical Gravitational Lensing Experiment data. A similar task of searching anomalous periodic variables was solved by  Chan et al. \cite{chan2022} in ZTF~DR2. Aleo et al.  \cite{aleo2022} used simulated light curves to search for counterparts in ZTF~DR4, identifying 11 non-catalogued transients.

However, applying AD methods to real astronomical data immediately raises a conceptual difficulty. In a statistical sense, an outlier is simply an object that does not fit the dominant distribution of the sample. 
In an astrophysical sense, an anomaly is an object that is scientifically interesting due to its unusual physical origin, manifested through an unusual observational appearance.
These two notions do not always coincide. In practice, statistically extreme objects are often dominated by image artefacts, processing errors, calibration issues, moving objects, or other forms of contamination (e.g., \cite{2021AJ....162..206S,2021MNRAS.502.5147M}). A central challenge is therefore to design strategies capable of moving from purely statistical outliers toward astrophysically meaningful anomalies.

The SNAD\footnote{\url{https://snad.space/}} project was created in this context as a framework for anomaly detection in astronomical survey data that combines machine-learning methods with expert knowledge \citep{volnovasnad}. Its development has followed three guiding principles. First, priority is given to real data, since no simulation can fully reproduce the complexity of astronomical observations and a pipeline can only be considered robust once it has been tested on real data. Second, the project adopts an active approach, in which expert feedback is incorporated into the anomaly-detection loop, because the notion of an anomaly is inherently subjective and depends on the scientific context. Third, the entire chain is addressed within the same framework, from feature extraction and algorithmic development to the astrophysical analysis and follow-up of the selected candidates. A central element that allows experts to analyse anomaly candidates is the SNAD ZTF Viewer\footnote{\url{https://ztf.snad.space/}} \citep{2023PASP..135b4503M}, which provides light curves, FITS image cutouts, catalogue cross-matches, light-curve modelling, and links to external astronomical services for individual ZTF objects. This end-to-end strategy is essential, since AD becomes scientifically useful only when it leads to the identification and interpretation of
genuinely interesting objects.

In this proceeding, we discuss the key terminology used to describe anomalies in SNAD, introduce our active AD approach, and summarize our recent results.

\section{Terminology}

There is no consensus in the astronomical literature on the definitions of terms such as anomalies, novelties, outliers, abnormal or unusual objects, rare phenomena, or unknown types. While these terms are often treated as synonyms (e.g., \cite{2022RAA....22h5006H}), some studies provide more explicit definitions, offering clear criteria for specific cases (e.g., \cite{2024ApJ...974..172A}, see Section 3).

In SNAD, we distinguish between machine-detected \textbf{outliers} and astrophysical \textbf{anomalies}, where the latter refers to objects or phenomena\footnote{The difference between object and phenomenon is essential. Examples of objects include galaxies, neutron stars, black holes, planets,  asteroids etc. In contrast, phenomena encompass events with a characteristic timescale, such as supernovae (SNe), gamma-ray bursts (GRBs), solar flares etc.} of scientific interest to researchers.

\subsection{Outlier}
By outlier we assume a statistical deviation within the data or, in the context of ML anomaly detection algorithms, any object with the high outlier score. The number of outliers depends on the chosen threshold. While these objects are unique or unusual from the perspective of the algorithm, they are not always of astrophysical interest, e.g. artifacts or misclassifications in considered catalogs. A good example could be a detection of a binary microlensing event in the Open Supernova Catalog~\cite{2019MNRAS.489.3591P}.

The causes of these outliers can vary: they may result from poor feature engineering, an inappropriate choice of AD algorithm, or the specificities of the considered dataset.

\subsection{Anomaly}
\label{anomaly_def}
In the context of SNAD, the concept of anomaly includes the following categories:

\begin{itemize}
    \item \textbf{Unknown-unknowns} are completely unforeseen phenomena or objects that are not predicted by any current theory. While their discovery might be driven by the search for anomalies, the exact nature of what is being sought remains entirely unknown. Such findings could range from requiring minor extensions to existing theories to the development of entirely new theoretical models, possibly even introducing new physics.
    A good historical example is the discovery of GRBs in the late 1960s by the U.S. Vela satellites, which were originally designed to detect gamma radiation from potential secret nuclear tests in space \cite{1973ApJ...182L..85K}.
    \item \textbf{Known-unknowns} are phenomena or objects that have been theoretically predicted but has not yet received observational confirmation, or whose confirmation remains unreliable. Examples include pair-instability SNe (candidates in  \cite{2022RNAAS...6..122P,2024A&A...683A.223S}), dark matter annihilation or decay (candidates discussed in \citep{2008Natur.456..362C,2009Natur.458..607A, 2012JCAP...07..054B,2012JCAP...08..007W, 2014ApJ...789...13B,2014PhRvL.113y1301B}), primordial black holes, wormholes, Dyson spheres etc.
    While some of these discoveries are expected to eventually be made as they arise naturally from our fundamental theories and represent critical tests of our current understanding of the Universe (e.g., the existence of cosmic neutrino background or primordial gravitational waves), their confirmation awaits improvements in observational precision and the development of new instruments and experiments. Others, however, may come as complete surprises and fundamentally change our worldview (e.g., Dyson spheres~\cite{1960Sci...131.1667D}).
    \item \textbf{Rare-knowns} refer to rare types of astronomical objects or phenomena that have already been observed (e.g., kilonovae, repeating fast radio bursts, superluminous supernovae (SLSNe)). The rarity of such anomalies may arise from their intrinsically small numbers in nature or from observational constraints, including selection effects. In the latter case, they should only be considered as anomalies while the number of identified members remains significantly lower compared to other known astronomical objects and phenomena.
\end{itemize}

\underline{Notes:}

While belonging to a rare class (e.g., kilonovae, of which only approximately a dozen have been confidently confirmed; see references in \cite{2023ApJ...942...88Z}) is obvious to any astrophysicist, subtler anomalies may involve unusual behavior within a common class. For example, peculiar modulation cycles of some RR Lyrae stars \cite{2023A&A...678A.104M}, the appearance of anomalous features in a lensing light curve \cite{2024A&A...686A.234H}, or spectroscopically peculiar SNe \cite{2021ApJ...907L..18D}. Such anomalies are more likely to be identified by domain experts. For example, a specialist in SNe may overlook an interesting feature in a Cepheid light curve, while it would be immediately recognized by a Cepheid expert. This makes the definition of an anomaly \textbf{expert-dependent}, as it varies based on the expert’s knowledge and field of expertise.

Secondly, the definition of an anomaly depends on the goals of the project. Even though M-dwarf flares are well-known phenomena, identifying them in a specific dataset can still meet the criteria for rarity in that context (e.g., \cite{2024MNRAS.533.4309V}). Another example is the study by \cite{2024ApJ...974..172A}, where spectroscopic anomalies are defined as any transient whose spectroscopic label is not of Type Ia-normal, Ia-91T-
like, II-normal, or IIP supernovae. This makes the definition of anomaly \textbf{task-dependent}, as it is tied to the specific scientific objectives or the dataset under consideration.

\section{Active approach}
The goal of ``healthy'' anomaly detection is \textit{to ensure that a large fraction of outliers are anomalies}. To achieve this, SNAD employs an \textbf{active} anomaly detection approach. 

Active anomaly detection algorithms adapt to human expert preferences during the learning process. These algorithms iteratively refine their models based on expert feedback, enabling them to better identify objects of scientific interest. 

To support this approach, we developed \texttt{Coniferest}\footnote{\url{https://coniferest.snad.space/}}~\cite{2024arXiv241017142K}, a Python library which offers implementations of three anomaly detection algorithms: the Isolation Forest and two active learning algorithms with an ``expert in loop'' --- Active Anomaly Discovery (AAD, \cite{Das2017, 2021A&A...650A.195I}) and PineForest, the latter being a method developed by our group (Korolev et al., in prep.).

\section{Projects and main results}
As stated in Section~\ref{anomaly_def}, the definition of an anomaly is both expert- and task-dependent. As a result, several sub-projects have emerged within SNAD, each focusing on different aspects of anomaly detection. 

\subsection{Traditional anomaly detection}
By applying different AD algorithms to the photometric data from the Open Supernova Catalog (OSC, \cite{2017ApJ...835...64G}) and the Zwicky Transient Facility survey \citep{Bellm_2019}, we identified a few dozen potentially interesting anomalies, as described in \cite{2019MNRAS.489.3591P}, \cite{2021MNRAS.502.5147M}, and \cite{2023RNAAS...7..155V}.

Among the diverse findings, there are candidates in SNe (e.g., \texttt{oid\footnote{ZTF Data Release object identifier; \url{https://ztf.snad.space/}}} = \texttt{795209200003484}), pre-main-sequence stars (e.g., \texttt{oid} = \texttt{807210200026027}), novae (e.g., \texttt{oid} = \texttt{695211100022045}), active galactic nuclei (AGNs; e.g., \texttt{oid} =  \texttt{718211400012193}), luminous blue variables (e.g., \texttt{oid} = \texttt{695211100131796}), a binary microlensing event AT~2021uey (\texttt{oid} = \texttt{643210400013909}), a radio source NVSS J080730+755017 with an optical counterpart (\texttt{oid} = \texttt{858205100001741}), and a RS Canum Venaticorum star (\texttt{oid} = \texttt{695211200019653}).

Furthermore, we also discovered numerous new variable stars (e.g., \texttt{oid} = \texttt{807212100012737}) and identified several cases of misclassification in the OSC, such as SN~2006kg, which is an AGN, and Gaia16aye, a binary microlensing event misclassified as supernova.

\subsection{Supernovae as anomalies}
This study represents the first science case testing the SNAD active learning pipeline~\cite{2023A&A...672A.111P}.

We analyzed 70 ZTF fields at high galactic latitudes, comprising $\sim$26.5 million light curves, and visually inspected 2100 outliers identified by the AAD algorithm. This resulted in 104 SN-like objects, 57 of which were newly reported to the Transient Name Server (TNS), while 47 had already been mentioned in other catalogs as either confirmed SNe or candidates. The light curves of unreported transients were visually inspected and fitted with supernova models, assigning to them a probable photometric classification: Ia, Ib/c, IIP, IIL, or IIn. Additionally, we identified slow-evolving transients as promising SLSNe, along with a few other non-catalogued events such as M-dwarf flares and AGNs.

This work demonstrates the effectiveness of the human-machine interaction underlying the AAD strategy. The SNe discovered with AAD, along with those identified by the SNAD Transient Miner \cite{2022NewA...9601846A}, form the basis of the SNAD catalog of transients: \url{https://snad.space/catalog/}.

\subsection{Superluminous supernovae as anomalies}
Superluminous supernovae are transients significantly brighter ($L_{\rm max} \gtrsim 10^{44}$ erg~s$^{-1}$) than ordinary supernovae, representing a relatively rare class of objects~\footnote{As of January 2025, TNS lists 278 classified SLSNe. Also, a catalog specifically focused on SLSNe~I includes 262 events~\cite{2024MNRAS.535..471G}.}~\cite{2024arXiv240712302M}. The exact mechanisms of SLSN explosions, which explain their observational properties, remain unknown. 
The explanations of their luminosity include the production of large amounts of radioactive $^{56}$Ni (exceeding several solar masses, sometimes even dozens), interaction of the ejecta with a dense circumstellar medium (CSM), magnetar-powered energy deposition, and fallback accretion onto a newly formed black hole (for a review, see \cite{2019ARA&A..57..305G}). SLSNe can be detected at high redshifts ($z\geq 2$), making them valuable for studying star formation in the early Universe, the interstellar medium, and even for cosmological applications \cite{2018SSRv..214...59M}.

The first SNAD SLSN candidates were identified during a supernova search using active learning applied to ZTF DR3 photometric data \cite{2023A&A...672A.111P}. One of the candidates, SNAD160, turned out to be a possible pair-instability SN, as its light curve exhibits a slow rise and decay consistent with predictions for such events~\cite{2022RNAAS...6..122P}.

In a follow-up study by \cite{2024arXiv241021077M}, the PineForest active AD algorithm was used to search for SLSNe in ZTF data. Starting with a dataset of $\sim$14 million objects and using 8 previously confirmed SLSN light curves as priors, we scrutinized 120 candidates. This led to the identification of 8 SLSN candidates, 2 of which (AT 2018moa and AT 2018mob) had not been reported previously.

\subsection{M-dwarf flares as anomalies}
In this project, we focus on searching for M-dwarf flares in ZTF data. M-dwarfs are main candidates for hosting extraterrestrial life, making it crucial to understand their activity. Moreover, such phenomena provide valuable insights into the magnetic activity of astronomical objects, including the Sun.

In \cite{2024MNRAS.533.4309V} we analyzed over 35 million ZTF light curves and constructed the largest catalogue of ZTF flaring stars to date, identifying 134 flares with amplitudes ranging from $-$0.2~mag to $-$4.6~mag, including repeated flares.

\subsection{Photometric artefacts as anomalies}
A significant portion of the outliers identified by AD algorithms consists of artefacts. For example, 68 percent of all outliers analyzed in \cite{2021MNRAS.502.5147M} were found to be bogus light curves. Even with an active learning approach, it is impossible to reject all artefacts immediately, as they are highly diverse and occupy various regions of the parameter space. 

During the expert analysis of outliers, we created the SNAD Knowledge Database~\cite{2023PASP..135b4503M}, which contains a few thousand labeled ZTF objects, approximately half of which are artefacts. We also stored the most interesting artefacts in the SNAD artefact catalog\footnote{\url{https://snad.space/art/}}, initially without specific scientific goals. Using this small but well-labeled sample, we developed a real-bogus classifier and integrated its results into the active anomaly detection pipeline as an additional feature. This significantly reduced the number of artefacts among outliers in regions of the feature space densely populated with artefacts~\cite{2025A&C....5100919S}.

Recognizing the utility of this approach, we started a dedicated campaign to systematically search for artefacts in ZTF using the PineForest algorithm -- where artefacts were treated as anomalies. The resulting artefact dataset includes 1127 FITS images of two sizes (28$\times$28 and 63$\times$63 pixels), along with a corresponding set of images of nominal objects. This was performed to support machine learning applications in astronomy, offering a real human-labeled dataset for the development and testing of algorithms (Sreejith et al., in preparation).

\section{Conclusions}

\begin{itemize}
    \item \textbf{Well-defined concepts are a fundamental part of scientific progress.} In SNAD, we distinguish machine-detected outliers from astrophysical anomalies and classify anomalies into unknown-unknowns, known-unknowns, and rare-knowns. Their interpretation is inherently expert-dependent and task-dependent, emphasizing the role of context in anomaly detection.
    \item \textbf{Active anomaly detection is an efficient approach} that significantly improves the identification of scientifically interesting objects and phenomena by incorporating expert feedback into the learning process.
    \item \textbf{The SNAD transient catalog has been created}, with some supernova candidates missed in the official ZTF alert stream. These findings can help avoid similar losses in future large-scale astronomical surveys. 
    \item \textbf{Several interesting anomalies were identified}, including SLSN candidates, one PISN candidate (SNAD160), binary microlensing events, previously uncatalogued AGNs, cataclysmic variables, and other rare variable stars.
    \item \textbf{The largest catalog of ZTF flaring stars to date was compiled}, representing 134 M-dwarf flares with amplitudes ranging from $-$0.2~mag to $-$4.6~mag.
    \item \textbf{Outliers in real data are often artifacts}, which can either be studied as a separate class or effectively filtered out to improve anomaly detection efficiency.
    \item \textbf{Anomaly detection systems must be built for experts}, ensuring interpretability and adaptability to domain-specific needs.
\end{itemize}

\section*{Acknowledgements}
We acknowledge Patrick Aleo, Boris Demkov, Vadim Krushinskiy, Anastasia Malancheva, Alexandra Novinskaya and  Anastasia Voloshina for their contribution to SNAD. M.~Kornilov, A.~Lavrukhina, M.~Pruzhinskaya, T.~Semenikhin acknowledge support from a Russian Science Foundation grant 24-22-00233, https://rscf.ru/en/project/24-22-00233/. Support was provided by Schmidt Sciences,~LLC. for K.~Malanchev.

\bibliographystyle{plain}
\bibliography{biblio}

\bigskip
\bigskip
\noindent {\bf DISCUSSION}

\bigskip
\noindent {\bf MARIA DAINOTTI:} Do you plan to have a citizen science project so that you can have on independent way to identify the bogus?

\bigskip
\noindent {\bf MARIA PRUZHINSKAYA:} Currently, we do not have plans for a citizen science project, but it is certainly an interesting possibility to consider for the future.

\bigskip
\noindent {\bf CLAUDIA MARASTON:} Do you have an estimate of the metallicity of the progenitor star of the PISN you detected? Can you extend your approach to a search for the most massive stars?

\bigskip
\noindent {\bf MARIA PRUZHINSKAYA:} We obtained the spectrum of the host of SNAD160 and estimated its metallicity. So far, our anomaly detection pipeline has not identified any other PISN candidates. 
In principle, active anomaly detection can be applicable to the search for any type of object, as long as they share common features present in the considered dataset.

\bigskip
\noindent {\bf INGYIN ZAW:} Description of how filtering out uninteresting candidates (to humans) that the machine thinks are interesting is clear. How do you deal with the possibility that the machine is missing candidates that humans would have found interesting?

\bigskip
\noindent {\bf MARIA PRUZHINSKAYA:} It is possible that the machine learning algorithm is missing really interesting objects, as there are many types of astrophysical phenomena. However, we believe that while truly anomalous objects are rare, they are not singular occurrences -- there are multiple anomalies waiting to be discovered. At this stage, we prioritize identifying unexpected and scientifically valuable cases rather than ensuring completeness. Finding even a single anomaly can be a significant breakthrough.

\end{document}